\documentclass[prx,twocolumn,amsmath,amssymb]{revtex4}

\usepackage[dvipdfmx]{xcolor}
\usepackage{bm}
\usepackage{dcolumn}
\usepackage[pdftex]{graphicx}    

\usepackage{slashed}
\usepackage{amsmath}\usepackage{accents}
\usepackage{physics}
\usepackage{mathbbol}\usepackage{bbm}
\usepackage{xpatch}

\newcommand{\1}{\mbox{1}\hspace{-0.25em}\mbox{l}}

\begin{document}
\NewDocumentCommand\zero{o o}{
    \scalebox{0.8}[1.5]{\kern .1mm$\mathbbm{o}$}%
    \IfValueT{#1}{%
        \IfNoValueT{#2}{_{#1}}
        \IfValueT{#2}{_{#1\times#2}}
    }
}
\NewDocumentCommand\eins{o o}{
    \mathbbm{1}%
    \IfValueT{#1}{%
        \IfNoValueT{#2}{_{#1}}
        \IfValueT{#2}{_{#1\times#2}}
    }
}

\title{
A formally exact real-space representation of the Berry phase on infinite lattices:
Applications to dipole and quadrupole moments
}

\author{Yusuke Onaya, Fumina Hamano, and Takahiro Fukui}
\affiliation{Department of Physics, Ibaraki University, Mito 310-8512, Japan}

\date{\today}

\begin{abstract}
Inspired by Kitaev's real-space representation of Chern numbers,  
we develop a real-space formulation of the Berry phase for infinite lattices.  
While the well-known Resta formula for the Berry phase is defined under periodic boundary conditions for finite lattices,  
our approach constructs the Berry phase directly on an infinite lattice  
without requiring momentum-space discretization.  
We apply this method to several disordered models to examine its validity.  
Furthermore, we attempt to generalize the real-space representation to the quadrupole moment,  
drawing an analogy to the generalization of the Resta formula for the quadrupole moment.  
\end{abstract}

\pacs{
}

\maketitle

\section{Introduction}
Topological invariants play a central role in various fields of modern physics,  
providing a fundamental framework for understanding the properties of matter  
\cite{Kane:2005aa,Kane:2005ab,Qi:2008aa,Schnyder:2008aa,Kitaev:2009fk}.  
Many of these invariants can be formulated in terms of the Berry connection,  
which serves as a foundation for describing topological properties in quantum systems.  
In condensed matter physics, the Berry connection is typically formulated in momentum space.  
However, this formulation inherently relies on translational symmetry  
and is less suited for systems with disorder, strong interactions, or spatial inhomogeneities.  
In such cases, a real-space representation becomes essential,  
offering a direct way to evaluate topological invariants via local physical quantities  
and providing a more intuitive perspective on topological properties.  

The Berry phase is given by a line integral of the Berry connection over momentum,  
which generally encodes geometric properties of quantum systems.  
In certain symmetry-protected topological phases, the Berry phase is quantized,  
serving as a one-dimensional topological invariant.  
As a real-space counterpart, the Berry phase is known as the Zak phase \cite{Zak:1989fk},  
which plays a fundamental role in the modern theory of electrical polarization in solids  
\cite{King-Smith:1993aa,Vanderbilt:1993fk}.  
A widely used formulation for computing polarization, both in momentum space and real space,  
is provided by Resta's approach \cite{Resta:1998aa}.  
This method has been extensively employed to evaluate the Berry phase in various contexts.  
However, the Resta formula is derived under periodic boundary conditions for finite lattices,  
where momentum is discretized, and its validity is strictly ensured only in the thermodynamic limit.  

In this paper, inspired by Kitaev's real-space representation of Chern numbers \cite{Kitaev:2006yg},  
we develop a real-space formulation of the Berry phase that is defined directly on an infinite lattice.  
While the Resta formula is well-defined under periodic boundary conditions,  
our approach does not rely on momentum-space discretization.  
We apply this framework to various disordered models, demonstrating its consistency with established methods.  
Furthermore, we extend this formulation to the quadrupole moment,  
drawing an analogy to the generalization of the Resta formula for topological-quadrupole systems.  
Our approach not only complements existing methodologies  
but also offers new perspectives on the bulk-edge correspondence,  
even in systems that are not necessarily topological.  

This paper is organized as follows.  
In Sec.~\ref{s:berrycon}, we reformulate the Berry connection using the square root operator  
of the spectral projector, which naturally leads to the real-space formulation of the Berry phase,  
as well as its extension to the quadrupole moment in Sec.~\ref{s:berryreal}.  
In Sec.~\ref{s:disorder}, we apply this formulation to various disordered systems,  
including a one-dimensional topological insulator, a $(1+1)$-dimensional pump system,  
and a two-dimensional higher-order topological insulator.  
Finally, in Sec.~\ref{s:summary}, we summarize our findings and discuss potential future applications.

\section{Berry connection}\label{s:berrycon}

In this section, we focus on systems with translational invariance  
and reformulate the Berry connection in momentum space.  
A key role in this reformulation is played by the square root operator  
of the spectral projector onto the ground state.  
This approach allows us to derive a real-space representation  
analogous to Kitaev's method \cite{Kitaev:2006yg}.

Let us start with the following Schr\"odinger equation in momentum space,
\begin{alignat}1
{\cal H}_k\psi_{nk}=\varepsilon_{nk}\psi_{nk},
\end{alignat}
where ${\cal H}_k$ and $\psi_{nk}$ represent the Bloch Hamiltonian and the Bloch state
with the crystal momentum $k$ and the band index $n$.
The Bloch wave function $\psi_{nk}$ is assumed to be normalized,
\begin{alignat}1
\psi_{nk}^\dagger\psi_{n'k}=\delta_{nn'}.
\end{alignat}
Introduce $M$ ground state multiplet  wave functions $\Psi_k\equiv(\psi_{1k},\psi_{2k},\cdots,\psi_{Mk})$.
Then, the matrix-valued Berry connection one-form is defined as
\begin{alignat}1
A\equiv \Psi^\dagger_k d\Psi_k=A_\mu dk_\mu,
\label{BerConDef}
\end{alignat}
where $A_\mu=\Psi_k^\dagger\partial_\mu \Psi_k$ with $\partial_\mu\equiv\partial_{k_\mu}$.

\subsection{Reformulation using the square root of the projector}

We define a symmetric matrix $Q$ as
\begin{alignat}1
Q_k &= \Psi_k\Psi_k^t = \sum_{n=1}^{M} \psi_{nk} \psi_{nk}^t,
\nonumber\\
&(Q^t = Q,\, Q^\dagger = Q^*),
\label{DefQk}
\end{alignat}
where $\psi^t$ denotes the transpose of $\psi$.  
In this subsection, we often suppress the $k$-dependence of $\psi_k$, $\Psi_k$, and $Q_k$ for simplicity.  
The matrix $Q$ plays a central role in the real-space representation of the Berry phase in the next section.  

The matrix $Q$ is related to the projector $P \equiv \Psi\Psi^\dagger$ via the relations  
\begin{alignat}1
QQ^\dagger = P, \quad Q^\dagger Q = P^*.
\label{QandP}
\end{alignat}
This implies that $Q$ can be regarded as a square root of the projector $P$.  
Further relationships hold:  
\begin{alignat}1
PQ = Q, \quad QP^* = Q.
\label{PandQ}
\end{alignat}

Now, let us consider the following one-form of $Q$:
\begin{alignat}1
\Psi^\dagger dQ \Psi^*
= \Psi^\dagger d\Psi + d\Psi^t \Psi^*.
\end{alignat}
Taking the trace of this equation leads to  
\begin{alignat}1
\tr Q^\dagger dQ = 2\tr A,
\end{alignat}
where $A$ is the Berry connection defined in Eq.~(\ref{BerConDef}),  
and $\tr$ denotes the trace over the occupied band indices, i.e., $\tr A = \sum_{n=1}^{M} A_{nn}$.  
Thus, we have established the relationship  
\begin{alignat}1
\tr A = \frac{1}{2} \tr Q^\dagger dQ.
\end{alignat}
This result shows that the Berry connection can be expressed in terms of the matrix $Q$,
which makes it easier to switch to a real-space representation.

\subsection{Gauge transformation}

While $P$ is gauge-invariant, $Q$ is gauge-dependent, as expected for the Berry connection.  
Under a gauge transformation,  
\begin{alignat}1
\Psi_k \rightarrow \Psi_k V_k,
\end{alignat}
where $V^\dagger V = \1$, the matrix $Q_k$ transforms as  
\begin{alignat}1
Q = \Psi\Psi^t \rightarrow \Psi V V^t \Psi^t.
\end{alignat}
Then, we obtain  
\begin{alignat}1
\tr Q^\dagger dQ \rightarrow \tr Q^\dagger dQ + 2\tr V^\dagger dV.
\end{alignat}
This ensures the correct gauge transformation property of the Berry connection,  
$\tr A \rightarrow \tr A + \tr V^\dagger dV$.

\section{Real-space representation of the Berry phase}\label{s:berryreal}

It is well known that the Chern number can be expressed in terms of the projector $P$,  
which allows for a real-space representation of the Chern number \cite{Kitaev:2006yg,PhysRevB.84.241106}.  
Similarly, the real-space representation of $Q$ naturally leads to a real-space representation of the Berry phase,  
as we will show in this section.

\subsection{Real-space representation of $Q$ and $P$}

The site-dependent form of the Bloch state is given by  
\begin{alignat}1
\psi_{j, nk} \equiv e^{ikj} \psi_{nk}, 
\label{BloSta}
\end{alignat}
where $j$ specifies the site index in a one-dimensional system.  
In what follows, we treat the Bloch states as matrices labeled by two indices:  
the site index $j$ and the quantum number index $nk$.  
The Hermitian conjugate of the Bloch states is formally denoted as $\psi^\dagger_{nk,j}$.  
The orthonormal relation is given by  
\begin{alignat}1
\sum_j \psi_{nk,j}^\dagger \psi_{j,n'k'} 
= 2\pi \delta_{nn'} \delta(k'-k).
\end{alignat}
In Eq.~(\ref{DefQk}), we defined the $k$-resolved matrix $Q_k$.  
To obtain a real-space representation,  
we treat the quantum numbers $n$ and $k$ on an equal footing  
and sum not only over $n$ but also over $k$ in Eq.~(\ref{DefQk}).  
Namely, using the Bloch state in Eq.~(\ref{BloSta}), we define  
\begin{alignat}1
Q_{jl} &\equiv
\int_{-\pi}^\pi \frac{dk}{2\pi} \sum_{n=1}^{M} \psi_{j,nk} \psi_{nk,l}^t
= \int_{-\pi}^\pi \frac{dk}{2\pi} e^{ik(j+l)} Q_k.
\label{QRel}
\end{alignat}
This expression shows that, due to translational symmetry,  
$Q_{jl}$ depends on $j$ and $l$ through their sum, i.e., $Q_{jl} = Q_{j+l}$.  
The inverse relation is given by  
\begin{alignat}1
Q_k = \sum_j e^{-ikj} Q_j.
\label{Qkjl}
\end{alignat}

In passing, let us also mention the real-space representation of the projector $P$
\begin{alignat}1
P_{jl} \equiv \int_{-\pi}^\pi \frac{dk}{2\pi} \sum_{n=1}^{M} \psi_{j,nk} \psi_{nk,l}^\dagger
= \int_{-\pi}^\pi \frac{dk}{2\pi} e^{ik(j-l)} P_k.
\label{PRea}
\end{alignat}
Due to translational invariance, $P_{jl}$ depends on $j$ and $l$ through their difference,  
leading to $P_{jl} = P_{j-l}$, as expected.  
Using these definitions, it is straightforward to show that the same relationships  
as in Eq.~(\ref{QandP}) hold. For instance,  
\begin{alignat}1
(QQ^\dagger)_{il} &= \sum_j Q_{ij} Q_{jl}^\dagger = P_{il}.
\end{alignat}

\subsection{Berry phase and polarization}

Now let us rewrite the Berry phase defined by
\begin{alignat}1
p&\equiv\frac{1}{2\pi i}\int_{-\pi}^\pi dk \tr A_k
\label{BerPhaMom}\\   
&=\frac{1}{4\pi i}\int_{-\pi}^\pi dk \tr Q_k^\dagger\partial_kQ_k.
\end{alignat}
Here, the Berry phase associated with the Bloch states is well-known in solid-state physics as representing polarization, which is why $p$ has been chosen as the notation.
Inserting Eq. (\ref{Qkjl}), the above becomes
\begin{alignat}1
p&=-\frac{1}{2}\sum_j\tr Q_j^\dagger jQ_{j}.
\label{BerRel}
\end{alignat}
Translational invariance enables rewriting the sum over $j$ as the product of matrices: Namely, setting 
$j\rightarrow j+i$, where $i$ is a certain fixed site, the above sum over $j$ can be written formally as 
\begin{alignat}1
p&=-\frac{1}{2}\sum_{j}\tr Q_{j+i}^\dagger(j+i)Q_{j+i}
\nonumber\\
&=-\frac{1}{2}\sum_{j}\tr Q_{ij}^\dagger (j+i)Q_{ji}
\nonumber\\
&=-\frac{1}{2}\tr (Q^\dagger\{X,Q\})_{ii},
\label{pQX}
\end{alignat}
where $X_{jl}=j\delta_{jl}$ is the position operator, and $\{A,B\}\equiv AB+BA$ is the anti-commutation relation.
The last equation tells that it does not depend on $i$ for systems with translational symmetry.
In the case of the Chern numbers, the projector $P$ instead of $Q$ yields a similar expression of the real-space
representation. However, it is because $P_{jl}$ depends on $j$ and $l$ through 
their difference $j-l$, as shown in Eq. (\ref{PRea}), 
that the Chern numbers are expressed  not by the anti-commutation relation but by the commutation relation
between $P$ and $X$ \cite{Kitaev:2006yg,PhysRevB.84.241106}.
The expression in Eq.~(\ref{pQX}), which involves the conventional position operator $X$, is still ill-defined for an extended system.
One way to avoid it is to consider the finite periodic systems with periodic boundary condition 
and use the exponentiated position operator \cite{PhysRevB.84.241106}. 
Another way is to introduce partitions in the real-space and  projectors onto them \cite{Kitaev:2006yg}.
In this paper, we adopt the latter approach.

To this end,  let us introduce the following operator,
\begin{alignat}1
\varSigma_{ij}\equiv \sigma_i\delta_{ij},\quad \sigma_i
=\frac{1}{2}\mbox{sgn}\,i \quad (\mbox{sgn}\,0\equiv 0).
\end{alignat}
This can be regarded as the projector used in \cite{Kitaev:2006yg}, but shifted by $-1/2$.
For generic matrices $A_{ij}$ and $B_{ij}$, we have
\begin{alignat}1
(A\{X,B\})_{ij}&=\sum_jA_{ik}(k+j)B_{kj},
\nonumber\\
(A\{\varSigma,B\})_{ij}&=\sum_jA_{ik}(\sigma_k+\sigma_j)B_{kj}.
\end{alignat}
Restricting our discussions to the translationally invariant matrices $A_{ij}=A_{i+j}$ and $B_{ij}=B_{i+j}$
such as $Q$ in Eq. (\ref{QRel}), we can show
\begin{alignat}1
\tr (A\{X,B\})_{ii}&=\sum_j\tr A_{i+j}(j+i)B_{j+i}
\nonumber\\
&=\sum_j\tr A_jjB_j.
\end{alignat}
We stress here again that this equation means that it does not depend on $i$. On the other hand, 
\begin{alignat}1
\Tr (A\{\varSigma,B\})&=\sum_{i,j}\tr A_{i+j}(\sigma_j+\sigma_i)B_{j+i}
\nonumber\\
&=\sum_j\tr A_j\sum_i(\sigma_{j-i}+\sigma_i)B_j
\nonumber\\
&=\sum_j\tr A_jjB_j,
\end{alignat}
where Tr means $\Tr A=\sum_i\tr A_{ii}$, and we have used the relationship,
$\sum_i(\sigma_{j-i}+\sigma_i)=j$.
Thus, we reach 
\begin{alignat}1
\tr(A\{X,B\})_{ii}=\Tr A\{\varSigma,B\}.
\end{alignat}
Applying to Eq. (\ref{pQX}), this equivalence allows the following real space representation,
\begin{alignat}1
p=-\frac{1}{2}\Tr Q^\dagger\{\varSigma,Q\}.
\label{ReaBerFinal}
\end{alignat}
This is formally exact expression of the polarization (Berry phase) on the infinite one-dimensional lattice.
Equation (\ref{ReaBerFinal}) includes two terms $Q^\dagger \varSigma Q$ and $Q^\dagger Q\varSigma$,
each of which is divergent, since the function $\sigma_j$ is constant up to $|j|\rightarrow\infty$.
 Therefore, the trace should be taken after computing the summation of these two matrices.

\subsection{Example: gSSH model}\label{s:pri_gSSH}

Before applying our formula to nontrivial models, we first verify its validity using simple topological models.  
Let us consider the generalized Su-Schrieffer-Heeger (gSSH) model with long-range hopping,  
defined as \cite{PhysRevB.89.224203,PhysRevB.103.224208,PhysRevB.110.045437}  
\begin{alignat}1
H_k &= \begin{pmatrix}  & \Delta_k^\dagger \\ \Delta_k & \end{pmatrix},  
\quad (\Delta_k \equiv t_1 + t_2 e^{ik} + t_3 e^{2ik}).
\label{gSSH}
\end{alignat}
Since this model possesses chiral symmetry,  
the gapped ground state at half-filling is classified by the winding number $w = 0, 1, 2$.  
A typical case in the atomic limit \cite{Ryu:2002fk} for each phase is given by  
$t_2 = t_3 = 0$ ($w=0$), $t_1 = t_3 = 0$ ($w=1$), and $t_1 = t_2 = 0$ ($w=2$),  
which are labeled as $(100)$, $(010)$, and $(001)$, respectively.  
The wave function of the lower band in each case is given by  
\begin{alignat}1
\psi_{k}^{(100)} &= \frac{1}{\sqrt{2}} \begin{pmatrix} 1 \\ -1 \end{pmatrix},
\nonumber\\
\psi_{k}^{(010)} &= \frac{1}{\sqrt{2}} \begin{pmatrix} 1 \\ -e^{ik} \end{pmatrix},
\nonumber\\
\psi_{k}^{(001)} &= \frac{1}{\sqrt{2}} \begin{pmatrix} 1 \\ -e^{2ik} \end{pmatrix}.
\label{SshWav}
\end{alignat}
It is straightforward to verify that the polarization in the momentum space in Eq.~(\ref{BerPhaMom})  
for these cases is directly given by  
\begin{alignat}1
p^{(100)} &= 0, \quad p^{(010)} = \frac{1}{2}, \quad p^{(001)} = 1.
\label{SshPol}
\end{alignat}
The normalization of the wave functions in Eq.~(\ref{SshWav}) leads to  
the relation between the polarization and the winding number, $p = w/2$.  

Next, we perform the real-space calculation using the matrix $Q$.  
From the wave functions in Eq.~(\ref{SshWav}), the matrix $Q_k$ in Eq.~(\ref{DefQk}) is computed as  
\begin{alignat}1
Q_k^{(100)} &= \frac{1}{2} \begin{pmatrix} 1 & -1 \\ -1 & 1 \end{pmatrix},
\nonumber\\
Q_k^{(010)} &= \frac{1}{2} \begin{pmatrix} 1 & -e^{ik} \\ -e^{ik} & e^{2ik} \end{pmatrix},
\nonumber\\
Q_k^{(001)} &= \frac{1}{2} \begin{pmatrix} 1 & -e^{2ik} \\ -e^{2ik} & e^{4ik} \end{pmatrix}.
\end{alignat}
Then, the real-space representation of $Q$ in Eq.~(\ref{QRel}) is obtained as  
\begin{alignat}1
Q_{jl}^{(100)} &=
\frac{1}{2} \begin{pmatrix} \delta_{j+l,0} & -\delta_{j+l,0} \\ -\delta_{j+l,0} & \delta_{j+l,0} \end{pmatrix},
\nonumber\\
Q_{jl}^{(010)} &=
\frac{1}{2} \begin{pmatrix} \delta_{j+l,0} & -\delta_{j+l+1,0} \\ -\delta_{j+l+1,0} & \delta_{j+l+2,0} \end{pmatrix},
\nonumber\\
Q_{jl}^{(001)} &=
\frac{1}{2} \begin{pmatrix} \delta_{j+l,0} & -\delta_{j+l+2,0} \\ -\delta_{j+l+2,0} & \delta_{j+l+4,0} \end{pmatrix}.
\end{alignat}
It is straightforward to compute the polarization via Eq.~(\ref{BerRel}),  
which reproduces the results in Eq.~(\ref{SshPol}).  

Finally, we verify the formula in Eq.~(\ref{ReaBerFinal}).  
Using its definition, the polarization can be rewritten as  
\begin{alignat}1
p &= \frac{-1}{2} \sum_{j,l,k} \tr (Q^\dagger_{jl} \varSigma_{lk} Q_{kj} + Q^\dagger_{jl} Q_{lk} \varSigma_{kj})
\nonumber\\
&= \frac{-1}{2} \sum_{j,l} (\sigma_l + \sigma_j) \tr Q^\dagger_{jl} Q_{lj}.
\label{pReMid}
\end{alignat}
For each case, we find  
\begin{alignat}1
\tr Q^{(100)\dagger}_{jl} Q_{lj}^{(100)} &= \delta_{j+l,0},
\nonumber\\
\tr Q^{(010)\dagger}_{jl} Q_{lj}^{(010)} &= \frac{1}{4} (\delta_{j+l,0} + 2\delta_{j+l+1,0} + \delta_{j+l+2,0}),
\nonumber\\
\tr Q^{(001)\dagger}_{jl} Q_{lj}^{(001)} &= \frac{1}{4} (\delta_{j+l,0} + 2\delta_{j+l+2,0} + \delta_{j+l+4,0}).
\end{alignat}
Summing over $l$ in Eq. (\ref{pReMid}) leads to  
\begin{alignat}1
p^{(100)}&=
\frac{-1}{2}\sum_{j}(\sigma_{-j}+\sigma_{j})=0,
\nonumber\\
p^{(010)}&=\frac{-1}{8}\sum_{j}
\big[\sigma_{-j}+\sigma_{j}+2(\sigma_{-j-1}+\sigma_{j})
\nonumber\\
&\qquad\qquad+(\sigma_{-j-2}+\sigma_{j})\big]=\frac{1}{2},
\nonumber\\
p^{(001)}&=
\frac{-1}{8}\sum_{j}
\big[\sigma_{-j}+\sigma_{j}+2(\sigma_{-j-2}+\sigma_{j})
\nonumber\\
&\qquad\qquad+(\sigma_{-j-4}+\sigma_{j})\big]=1,
\label{InfLat}
\end{alignat}
which correctly reproduces Eq.~(\ref{SshPol}).  

As demonstrated, the matrix $Q$ is computed on an infinite lattice,  
ultimately yielding the same results as in Eq.~(\ref{SshPol}).  
Thus, we confirm that the formula in Eq.~(\ref{ReaBerFinal}) correctly describes the SSH model.  
In Sec.~\ref{s:dssh}, we show that even in the presence of disorder,  
our formula provides consistent results for topological transitions,  
in agreement with the Resta formula.

\subsection{Calculation for finite systems}\label{s:finite}

In the previous subsection, we demonstrated that our formula correctly reproduces the polarization.  
There, the simple atomic limit allowed us to compute the matrix $Q$ exactly in real space.  
However, when applying our formula to more complex models, numerical calculations become inevitable.  

For winding numbers, as discussed by Kitaev, truncating infinite matrices to finite ones  
has proven to be highly effective \cite{Kitaev:2006yg}, and appropriate truncation sizes  
have been analyzed in Ref.~\cite{PhysRevB.110.045437}.  
In the present case, we argue that our formula remains valid even without truncation,  
as long as open boundary conditions are imposed.  
Before demonstrating this in the next section, we first derive a formula  
suitable for a finite-dimensional matrix $Q$.  

When the matrix $Q$ is of finite dimension,
we can utilize the conventional trace formula,  
which allows us to rewrite Eq.~(\ref{ReaBerFinal}) in a simpler form as  
\begin{alignat}1
p&=-\frac{1}{2}\Tr (QQ^\dagger+Q^\dagger Q)\varSigma=-\frac{1}{2}\Tr (P+P^*)\varSigma
\nonumber\\
&=-\Re \Tr P\varSigma,
\label{ReaBerFinite}
\end{alignat}
where $\Psi$ denotes the set of ground-state multiplet wave functions in real space.  
This equation represents one of the main results of this paper.  

Again, we emphasize that {\it this formula holds under open boundary conditions.}
Moreover, note that Eq.~(\ref{ReaBerFinite}) is {\it gauge-invariant,}  
whereas Eq.~(\ref{ReaBerFinal}) is gauge-dependent.  
Such behavior is commonly observed: the Berry connection in momentum space is inherently gauge-dependent,  
but once momentum is discretized and the Berry connection is defined as a link variable,  
it becomes gauge-invariant \cite{FHS05}.  
This property can also be understood in terms of the Resta formula.  

The formula in Eq.~(\ref{ReaBerFinite}) is simple enough to reveal its quantization,  
which arises due to symmetries such as chiral symmetry and inversion symmetry.  
In the following section, we examine how Eq.~(\ref{ReaBerFinite}) reproduces the polarization  
under open boundary conditions using the gSSH model in the atomic limit.

\subsubsection{Trivial phase}

We consider a system where unit cells, each consisting of two sublattices A and B,  
are labeled by $-n_-\le j\le n_+$.  
The total number of unit cells is given by $N = n_+ + n_- + 1$.  
The real-space Hamiltonian under open boundary conditions is given by  
\begin{alignat}1
H^{(100)} = t_1 \begin{pmatrix} & \1_N \\ \1_N & \end{pmatrix},
\end{alignat}
where the dimension $N$ of the identity matrix $\1$ has been explicitly indicated.  
The degenerate negative-energy eigenstates of this Hamiltonian are  
\begin{alignat}1
\Psi = \frac{1}{\sqrt{2}} \begin{pmatrix} \1_N \\ -\1_N \end{pmatrix}.
\end{alignat}
Thus, the polarization for a finite system, as given by Eq.~(\ref{ReaBerFinite}), becomes  
\begin{alignat}1
p^{(100)}
&= \Tr \varSigma = \frac{n_+ - n_-}{2}.
\end{alignat}

This result suggests that we should choose $n_- = n_+$.  
The result on an infinite lattice, Eq.~(\ref{InfLat}),  
can be understood as first setting $n_- = n_+$  
and then taking the limits $n_- \rightarrow \infty$ and $n_+ \rightarrow \infty$,  
analogous to the principal value prescription for divergent integrals.  
In the following calculations, we assume $n_- = n_+ \equiv n_0$.

\subsubsection{Topological phases}\label{s:topological} 

In the second case, the Hamiltonian in real space under open boundary conditions is
\begin{alignat}1
H^{(010)}=t_2\left(\begin{array}{c|cc|c}&&&0\\ \hline&&\1_{N-1}&\\ &\1_{N-1} &&\\ \hline 0&&&\end{array}\right).
\end{alignat}
Then, $(N-1)$ negative-energy states have eigenstates
\begin{alignat}1
\Psi=\frac{1}{\sqrt{2}}\left(\begin{array}{c}\bm0^t\\ \hline \1_{N-1}\\-\1_{N-1} \\\hline  \bm0^t\end{array}\right)
\end{alignat}
where $\bm 0^t=(0,\dots,0)$ represents the zero vector of dimension $N-1$.
In addition, two zero-energy edge states appear:
\begin{alignat}1
\psi_{\rm e}=\left(\begin{array}{c}1\\ \hline  \\  \\\hline0\end{array}\right),
\left(\begin{array}{c}0 \\ \hline   \\ \\ \hline1\end{array}\right).
\end{alignat}
To lift the degeneracy of these edge states  
and ensure that the half-filled ground state is well-defined even under open boundary conditions,  
we introduce an infinitesimal mass term.  
As a result, one of the two edge states becomes part of the ground-state multiplet.  
It is straightforward to see that, similar to the trivial case,  
the bulk states contribute zero to the polarization,  
whereas the edge state contributes $\pm 1/2$.  
Thus, the total polarization of the ground state is  
\begin{alignat}1
p^{(010)}=\pm1/2,
\end{alignat}
where the sign is determined by the sign of the infinitesimal mass term.  
Likewise, in the case of $(001)$, two of the four edge states contribute to the Berry phase, yielding  
\begin{alignat}1
p^{(001)}=\pm 1.
\end{alignat}

These results indicate that Eq.~(\ref{ReaBerFinite}) primarily characterizes the polarization  
in terms of the edge states.  
This stands in sharp contrast to, for example, the Resta formula \cite{Resta:1998aa},  
which is designed to determine the polarization under periodic boundary conditions.

\subsection{Quadrupole moment}

The Resta formula for the dipole moment has been extended to the quadrupole moment  
\cite{PhysRevB.100.245135,PhysRevB.100.245134,PhysRevLett.125.166801,PhysRevLett.128.127601}  
to characterize corner states in higher-order topological insulators  
\cite{Benalcazar:2017aa,Benalcazar:2017ab,Schindler:2018ab,Imhof:2018aa,  
Kunst:2018aa,Ezawa:2018aa,Schindler:2018aa,Fukui:2018aa,Hayashi:2018aa,Hashimoto:2017aa,Ota:2019aa,Zhang:2019ab}.  
Following this approach, we extend the real-space representation of the polarization (\ref{BerRel})  
to the quadrupole moment.  

Consider a two-dimensional lattice, where sites are labeled by $j = (j_x, j_y)$.  
A natural extension of Eq.~(\ref{BerRel}) to the quadrupole moment is given by  
\begin{alignat}1
q_{xy}=-\frac{1}{2}\sum_j \tr Q_j^\dagger j_xj_yQ_j.
\label{Qxy1}
\end{alignat}  
Correspondingly, we introduce an alternative expression
\begin{alignat}1
q_{xy}=-\frac{1}{2}\Tr Q^\dagger\{\varSigma_{xy},Q\},
\label{Qxy2}
\end{alignat}  
where $\varSigma_{xy}=\sigma_{j_x}\sigma_{j_y}\delta_{jl}$.  
We now show the equivalence of these two definitions.  
For translationally invariant systems, where $Q_{ij} = Q_{i+j}$, Eq.~(\ref{Qxy2}) yields  
\begin{alignat}1
q_{xy}&=-\frac{1}{2}\sum_{i,j}\tr Q^\dagger_{ij}(\sigma_{j_x}\sigma_{j_y}+\sigma_{i_x}\sigma_{i_y})Q_{ji}
\nonumber\\
&=-\frac{1}{2}\sum_{i,j}\tr Q^\dagger_{i+j}(\sigma_{j_x}\sigma_{j_y}+\sigma_{i_x}\sigma_{i_y})Q_{j+i}
\nonumber\\
&=-\frac{1}{2}\sum_{i,j}\tr Q^\dagger_{j}(\sigma_{j_x-i_x}\sigma_{j_y-i_y}+\sigma_{i_x}\sigma_{i_y})Q_{j}.
\end{alignat}  
Using the identity  
\begin{alignat}1
\sum_{i}(\sigma_{j_x-i_x}\sigma_{j_y-i_y}+\sigma_{i_x}\sigma_{i_y})=j_xj_y,
\end{alignat}  
we conclude that Eqs.~(\ref{Qxy1}) and (\ref{Qxy2}) are equivalent in translationally invariant systems.
This provides a straightforward generalization of the polarization formula (\ref{ReaBerFinal})  
to the quadrupole moment.  

For numerical calculations on finite-size systems with open boundary conditions,  
the above formula reduces to  
\begin{alignat}1
q_{xy}=-\Re\Tr P\varSigma_{xy},
\label{ReaQuaFinite}
\end{alignat}  
which corresponds to the polarization formula (\ref{ReaBerFinite}).

\section{Applications to disordered systems}\label{s:disorder}

So far, we have derived the formal expression for the polarization on an infinite lattice (\ref{ReaBerFinal}) 
and its reduced version (\ref{ReaBerFinite}), which is suitable for numerical computations on finite systems. 
The reduced formula has revealed that edge states play an important role, 
as demonstrated using simple models in the atomic limit. 
In this section, we examine how the reduced formula applies in more general settings, particularly in disordered systems.

\subsection{Disordered gSSH model: Quantized polarization due to symmetry protection}
\label{s:dssh}
The first example is the gSSH model, which has been studied in Secs.~\ref{s:pri_gSSH} and \ref{s:finite}.
Here, we introduce nontrivial disorder potentials while preserving chiral symmetry
and examine the resulting topological changes.
Although various real-space methods for computing winding numbers suitable for the gSSH model
have been developed
\cite{Resta:1998aa,Kitaev:2006yg,PhysRevB.89.224203,PhysRevB.103.224208,
PhysRevA.103.043310,PhysRevB.110.045437},
here we compute the polarization using our formula and compare the result with that obtained from the Resta formula.

To describe the gSSH model defined in Eq. (\ref{gSSH}) in real space, we consider a one-dimensional lattice consisting of unit cells, each composed of two species, denoted as $A$ and $B$. 
The general noninteracting Hamiltonian is then given by
\begin{alignat}1
H=\left(\begin{array}{cc}\bm c_{A}^\dagger,\bm c_{B}^\dagger\end{array}\right)
\left(\begin{array}{cc}\Gamma_A& \Delta^\dagger\\ \Delta & \Gamma_B\end{array}\right)
\left(\begin{array}{c}\bm c_A\\\bm c_B\end{array}\right),
\label{Ham}
\end{alignat}
where $\bm c_A^T=(\cdots,c_{A,-1},c_{A,0},c_{A,1},\cdots)$ represents 
the annihilation operator for species $A$,
and similaly for $\bm c_B$. 
The gSSH model under consideration is defined by the following specific matrix elements 
$\Gamma_{A,B}$ and $\Delta$, 
\begin{alignat}1
&\Gamma_{A,ij}=-\Gamma_{B,ij}=\epsilon \delta_{ij},
\nonumber\\
&\Delta_{ij}=t_{1,i}\delta_{ij}+t_{2,i}\delta_{i,j+1}+t_{3,i}\delta_{i,j+2},
\end{alignat}
where $\epsilon$  is an infinitesimal chiral symmetry-breaking term, as mentioned in Sec. \ref{s:topological},  
introduced to lift the hybridization of edge states localized at opposite ends.  
As studied in Secs. \ref{s:pri_gSSH} and \ref{s:finite}, this system exhibits three types of quantized polarization.  
Let us introduce hopping disorder into this system, which breaks translational symmetry, and calculate the polarization  in the real-space representation.  

\begin{figure}[htbp]
\begin{center}
\begin{tabular}{c}
\includegraphics[width=0.95\linewidth]{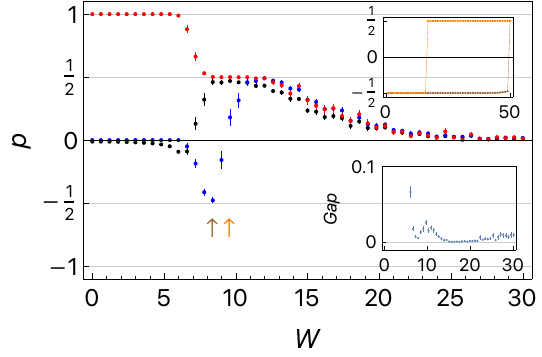}
\end{tabular}
\caption{
Real-space calculation of the polarization for the gSSH model.
The red, blue, and black dots denote the real-space polarization (\ref{ReaBerFinite})
under the open boundary condition, the Resta formula (Wilson-loop calculation) 
under the periodic boundary condition, and also the same Resta formula under the 
open boundary condition.
Each dot stands for the averaged value over 50 ensemble.
The system size is $n_0=100$ ($N=201$ site system).
The upper inset shows the distribution of polarization values across all disorder realizations at the strengths indicated by the arrows, and
the lower inset shows the energy gap at half-filling.
}
\label{f:gssh}
\end{center}
\end{figure}
To preserve chiral symmetry, we consider the following disordered hoppings:
\begin{alignat}1
t_{1,i}&=0+W\delta t_{1,i},
\nonumber\\
t_{2,i}&=1+\frac{W}{2}\delta t_{2,i},
\nonumber\\
t_{3,i}&=2,
\end{alignat}
where $ \delta t_{1,i} $ and $ \delta t_{2,i} $ are independent random real variables  
satisfying $ \delta t_{1,i}, \delta t_{2,i} \in [-0.5,0.5] $.  
This model has been studied in Refs.  
\cite{PhysRevB.89.224203,PhysRevB.103.224208,PhysRevB.110.045437}  
using the real-space representation of the winding numbers.

Fig. \ref{f:gssh} presents the polarization computed using various real-space representations.  
The red dots indicate the polarizations calculated using Eq. (\ref{ReaBerFinite}), the method proposed in this paper.  
For comparison, the polarization obtained via the conventional Resta formula,  
\begin{alignat}1
p_{\rm R} \equiv -\frac{1}{2\pi} \Im \log \det\left(\Psi^\dagger e^{i\frac{2\pi}{N} X} \Psi\right),
\quad \mbox{(mod } 1),
\label{Res}
\end{alignat}
is also shown, with blue and black dots representing the results under periodic and open boundary conditions, respectively.  
Here, $X$ denotes the position operator, as defined below Eq. (\ref{pQX}). 
To ensure a direct comparison with our method, we also compute the Resta formula under open boundary conditions, which is not its usual setting.

This model belongs to the BDI class and is thus classified by winding numbers rather than polarization.  
Previous studies \cite{PhysRevB.89.224203,PhysRevB.103.224208,PhysRevB.110.045437} have demonstrated that, in the absence of disorder, the model exhibits a winding number of $w = 2$ due to the presence of long-range hopping $t_2$.  
As disorder increases, the winding number decreases stepwise from 2 to 1 and eventually to 0.  
Indeed, the real-space calculation based on Eq. (\ref{ReaBerFinite}) accurately reproduces these winding numbers, including $w = 2$.  
This is because the present formula reflects the edge states, as discussed in Sec.~\ref{s:topological}.

On the other hand, the Resta formula (\ref{Res}) is manifestly defined modulo 1.  
Indeed, as seen in Fig.~\ref{f:gssh}, the two phases with winding numbers 2 and 0  
cannot be distinguished in numerical calculations, as the branch of the logarithm is determined by its principal values  
in our calculations.  
In contrast, the intermediate phase with winding number $w=1$ is characterized by a polarization of $1/2$.  
The Resta formula correctly reproduces this phase regardless of the boundary conditions,  
even though it is originally derived under the assumption of periodic boundary conditions.  
At first glance, the results in Fig.~\ref{f:gssh}, represented by blue dots,  
appear to fail to
capture the correct polarization values around $p=1/2$.  
However, this is because the polarization takes values of $\pm 1/2$,  
which are mathematically equivalent. When averaged, however, the result inevitably deviates from $\pm 1/2$.  
The upper inset shows all the ensembles of polarization at fixed disorder strengths.
This figure demonstrates that the point indicated by orange arrow indeed take values of $-1/2$ or $+1/2$,
confirming that the polarization is indeed $1/2 \mod 1$.  
Nevertheless, ensemble averaging leads to an apparent deviation from $\pm1/2$ in Fig.~\ref{f:gssh}.

\begin{figure*}[htb]
\begin{center}
\begin{tabular}{ccc}
\includegraphics[width=0.33\linewidth]{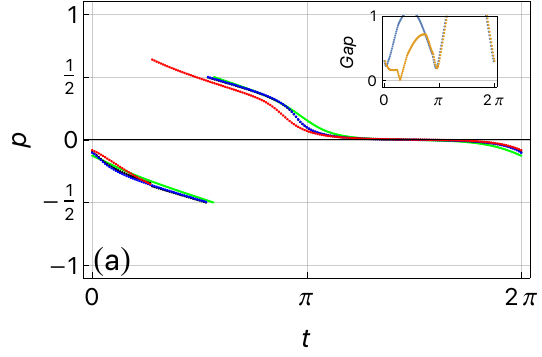}&
\includegraphics[width=0.33\linewidth]{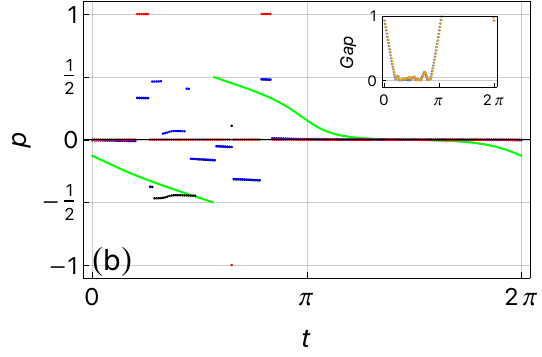}&
\includegraphics[width=0.33\linewidth]{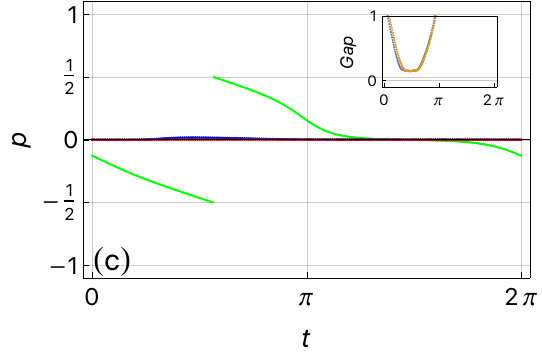}
\end{tabular}
\caption{
Examples of polarization as a function of $t$ for the disordered Rice-Mele model.  
Black and blue points represent the polarization computed using the Resta formula  
under open and periodic boundary conditions, respectively,  
while red points correspond to the polarization computed using the real-space formula (\ref{ReaBerFinite}).  
All calculations are performed for the same realization of disorder $\delta m_{I,i}$,  
with only the disorder strength $W$ varying.  
The parameters used are $t_0 = 1$, $t_d = 0.5$, $m_0 = 0.1$, and $m_d = 0.5$  
for a system of size $n_0 = 50$.  
Panels (a), (b), and (c) correspond to $W = 2$, $W = 10$, and $W = 17$, respectively,  
where the same disorder realization $\delta m_{i}$ is used in all cases.  
The shared green curve corresponds to the polarization of the clean system ($W = 0$).
The inset shows the zero-energy gap as a function of $t$,  
with light blue and pink points corresponding to periodic and open boundary conditions, respectively.  
}
\label{f:grm_bp}
\end{center}
\end{figure*}

\subsection{Rice-Mele model: Chern number in $1+1$-dimensional system}

The Rice-Mele model is a fundamental example of the Thouless pump \cite{Thouless:1983fk,Lohse:2016aa,Nakajima:2016aa}.
This model belongs to class A and is thus characterized by Chern numbers.
In the clean limit \cite{Wang:2013fk_pump}, Chern numbers are conventionally computed in momentum space
using the approach introduced in Ref.~\cite{FHS05}.
Various methods for computing Chern numbers in real space have been developed
\cite{Kitaev:2006yg,PhysRevB.80.125327,Prodan_2010,PhysRevB.84.241106,
Caio:2019aa,PhysRevA.100.023610,PhysRevLett.122.166602,PhysRevA.101.063606,
PhysRevB.103.155134,PhysRevA.103.043310,shiina2025realspacerepresentationsecondchern}.
In this subsection, we instead employ the polarization formula (\ref{ReaBerFinite}) to compute the Chern number.
This demonstrates that our method is applicable to generic systems, even in the absence of specific symmetry constraints.

We now consider the Hamiltonian (\ref{Ham}), which varies adiabatically with time $t$.
Its matrix elements are  explicitly given by
\begin{alignat}1
&\Gamma_{A,ij}=-\Gamma_{B,ij}=\gamma(t)\delta_{ij},  
\nonumber\\
&\Delta_{ij}=t_1(t)\delta_{ij}+t_2(t)\delta_{i,j+1}.  
\end{alignat}  
We assume that the parameters $\gamma$ and $t_{1,2}$ are $T$-periodic in time:
\begin{alignat}1
&\gamma(t)=m_0+m_d\cos (2\pi t/T),  
\nonumber\\
&t_{1}(t)=t_0- t_d\sin (2\pi t/T),  
\nonumber\\
&t_{2}(t)=t_0+ t_d\sin (2\pi t/T).  
\end{alignat} 

For this model, the Chern number is given by the change in polarization over one period.  
Specifically, we compute the polarization $p(t)$ at fixed $t$ and take the difference between $t = T$ and $t = 0$:  
\begin{alignat}1
c_1 = p(T) - p(0).
\label{Che}
\end{alignat}  
However, this expression is only formal, as the polarization $p(t)$ is defined on the principal branch of  
the logarithm in the case of the Resta formula, leading to discontinuous jumps as a function of $t$.  
In contrast, in our formulation, these discontinuities arise due to the crossing of edge states, which induces changes  
in the ground state configuration.  
As a result, it is necessary to evaluate $p(t)$ not only at $t = 0$ and $t = T$ but also at intermediate values of $t$  
to properly account for these discontinuities.

Let us introduce onsite disorder associated with $m_0$ in this model.  
Specifically, we add the following term $\delta\Gamma_{I}$ to $\Gamma_{I}$ ($I = A, B$):  
\begin{alignat}1
\delta\Gamma_{A,ij} =-\delta\Gamma_{B,ij} =W \delta m_{i} \delta_{ij},  
\end{alignat}  
where $\delta m_{i}$ are independent random real variables  
uniformly distributed in $[-0.5, 0.5]$.  

To numerically compute the Chern numbers via polarization,  
we first examine several examples of the polarization $p$ as a function of $t$,  
calculated for {\it a fixed realization of disorder $\delta m_{I,i}$}.  

We begin with the green curve in Fig.~\ref{f:grm_bp},  
which represents the polarization of the pure model in the absence of disorder ($W = 0$),  
computed using the Resta formula (\ref{Res}) in momentum space,  
or equivalently, via the Wilson loop method.  
This curve exhibits a jump from $p = -1/2$ to $p = 1/2$,  
which mathematically originates from the principal branch of the logarithm.  
If this artificial jump is connected smoothly,  
the total change from $t = 0$ to $t = T$ is $-1$,  
implying a Chern number of $-1$ via Eq.~(\ref{Che}).  

With this in mind, we now examine real-space calculations using  
Eq.~(\ref{ReaBerFinite}) as well as the Resta formula under both periodic and open boundary conditions,  
as shown by the dots in Fig.~\ref{f:grm_bp}.  
First, in the absence of disorder, all three methods --- our approach, the Resta formula with periodic boundary conditions,  
and the Resta formula with open boundary conditions --- perfectly agree with the polarization curve  
computed in momentum space, i.e., the green curve in Fig.~\ref{f:grm_bp}.

In the presence of weak disorder, as in panel (a), real-space computations yield a polarization profile  
that closely matches that of the clean model.  
Interestingly, in our method, the jump in polarization is slightly shifted:  
this shifted jump occurs at the point where the gap closes under open boundary conditions,  
as shown in the inset of Fig.~\ref{f:grm_bp}.
This suggests that our approach accurately captures the topological transitions associated with edge state crossings,  
distinguishing it from the Resta formula, which is influenced by the branch structure of the logarithm.  
Even under open boundary conditions, the Resta formula correctly reproduces the polarization,  
yielding the expected Chern number.  
As disorder strength increases, as in panel (b), the spectrum becomes gapless,  
and the polarization exhibits complex jumps due to multiple level crossings.  
In general, polarization jumps caused by level crossings between bulk states tend to occur in pairs,  
with upward and downward transitions canceling each other out,  
resulting in a net-zero contribution to the winding number.  
For sufficiently strong disorder, as in panel (c), a gap reopens,  
and the polarization becomes well-defined, yielding a zero winding number, i.e., a zero Chern number.

\begin{figure}[h]
\begin{center}
\begin{tabular}{c}
\includegraphics[width=0.95\linewidth]{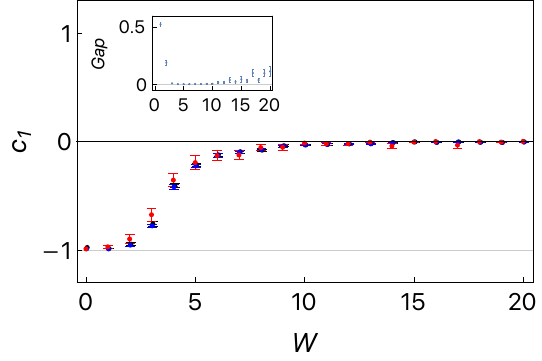}
\end{tabular}
\caption{
The Chern numbers, averaged over 50 ensembles,  as a function of the disorder strength $W$.  
The inset shows the minimum energy gap at zero energy.
}
\label{f:grm_ch}
\end{center}
\end{figure}

This observation suggests that computing the Chern number via polarization  
relies on identifying discontinuous jumps in $p(t)$.  
In practical numerical calculations, these jumps should be systematically detected and summed.  
The total accumulated jump then corresponds to the Chern number \cite{Hatsugai:2016aa}.  

In Fig.~\ref{f:grm_ch}, ensemble-averaged Chern numbers,  
determined from polarization jumps, are plotted as a function of disorder strength.  
The results indicate that the nontrivial pump phase remains stable up to $W \sim 2$,
beyond which the system enters a gapless regime  
before eventually transitioning into a trivial insulating phase as the bulk gap reopens.  
Importantly, the ensemble-averaged Chern number remains consistent across different calculation methods,  
confirming the robustness of our approach.

\begin{figure}[htb]
\begin{center}
\begin{tabular}{c}
\includegraphics[width=0.7\linewidth]{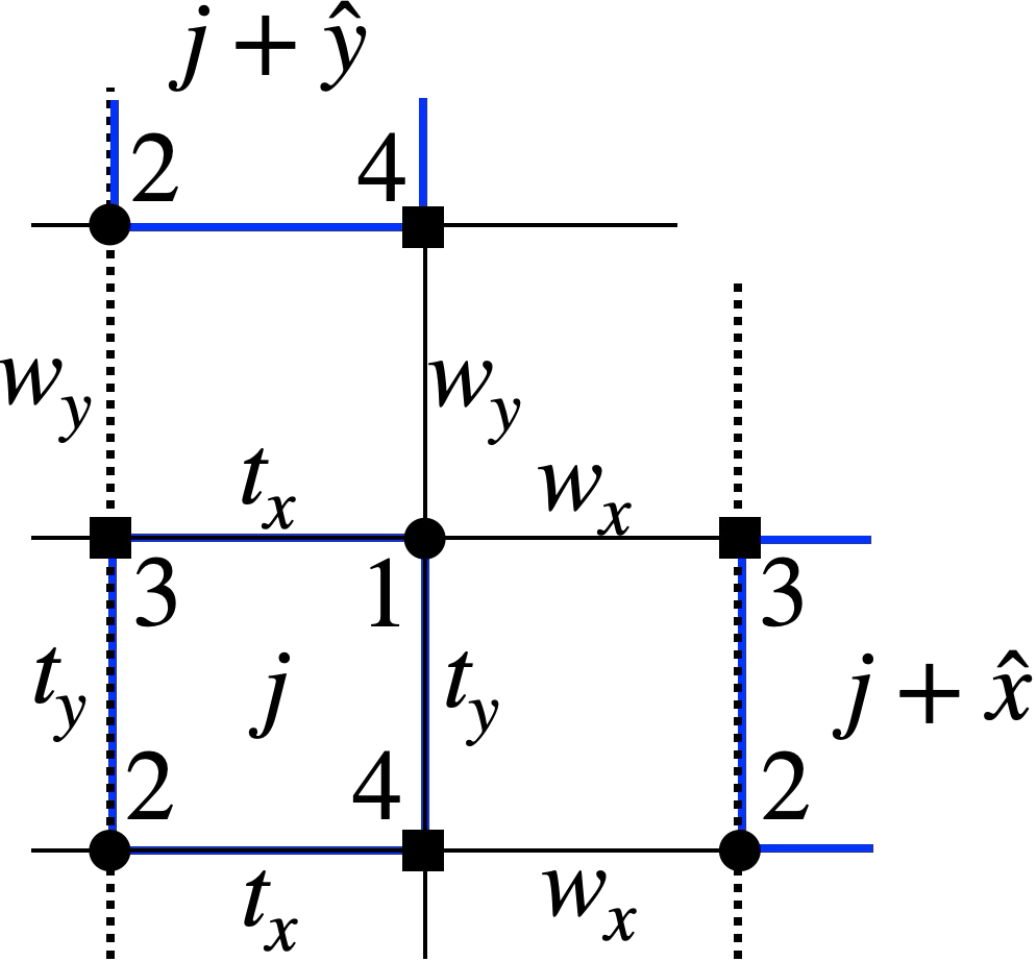}\\
\end{tabular}
\caption{
The square lattice with a $\pi$-flux per plaquette. Dotted lines stand for negative hoppings due to $\pi$-flux.
The unit cell, indicated by a blue square, consists of four sites.  
The $A$ and $B$ sublattices each  consists of two sites: sublattice $A$ includes sites 1 and 2 (represented by circles), while sublattice $B$ includes sites 3 and 4 (represented by squares).  
The unit vectors in the $x$- and $y$-directions are denoted by $\hat{x} = (1,0)$ and $\hat{y} = (0,1)$, respectively.
}
\label{f:bbh_lat}
\end{center}
\end{figure}

\subsection{BBH Model}

Finally, we examine how the quadrupole moment formula in Eq.~(\ref{ReaQuaFinite}) applies to disordered systems  
by comparing it with results obtained from the generalized Resta formula  
\cite{PhysRevB.100.245135,PhysRevB.100.245134,PhysRevLett.125.166801,PhysRevLett.128.127601}.  
The model discussed in this section was originally proposed by Benalcazar, Bernevig, and Hughes (BBH)  
to explore topological quadrupole insulators \cite{Benalcazar:2017aa,Benalcazar:2017ab}.  
Since then, it has been extensively studied as a prototype of higher-order topological insulators.  

\begin{figure}[htb]
\begin{center}
\begin{tabular}{c}
\includegraphics[width=0.95\linewidth]{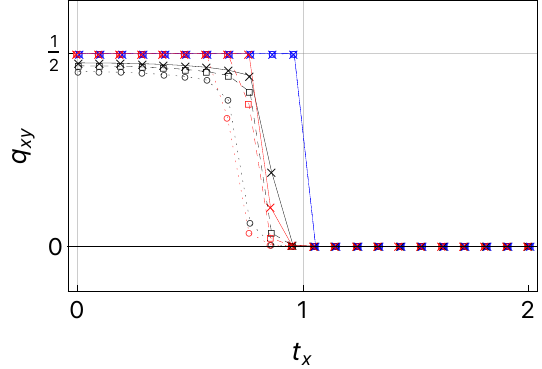}\\
\end{tabular}
\caption{
The quadrupole moment $q_{xy}$ of the clean BBH model  
as a function of $t_x$, with fixed parameters $t_y=0.1$ and $w_x=w_y=1$.  
Red, blue, and black points correspond to calculations using Eq.~(\ref{ReaQuaFinite}),  
the Resta formula in Eq.~(\ref{QuaRes}) with periodic boundary conditions,  
and the same Resta formula with open boundary conditions, respectively.  
Circles (connected by dotted lines), squares (connected by dashed lines),  
and crosses (connected by solid lines) represent system sizes of $n_0=10$, 15, and 20, respectively.  
}
\label{f:quad_fx}
\end{center}
\end{figure}

The BBH Hamiltonian is defined on a square lattice with a $\pi$-flux, as illustrated in Fig.~\ref{f:bbh_lat}.  
It is convenient to describe this two-dimensional model using shift operators.  
Let $j=(j_x,j_y)$ label a unit cell, which consists of a plaquette in the square lattice, and  
let $\bm c_{A(B)}$ in Eq.~(\ref{Ham}) be a suitably ordered two-dimensional fermion operator,  
where sublattice $A \,(B)$ includes sites labeled as $1,2 \,(3,4)$ in Fig.~\ref{f:bbh_lat}.  
We then introduce the shift operators $\delta_x$ and $\delta_y$, defined as  
$\delta_x c_{I,j} = c_{I,(j_x+1,j_y)}$ and $\delta_x^{-1} c_{I,j} = c_{I,(j_x-1,j_y)}$,  
with similar definitions for $\delta_y$, where $I=A1,A2,B3,B4$.
The BBH Hamiltonian is then given by Eq.~(\ref{Ham}) with  
\begin{alignat}{1}
&\Gamma_A=-\Gamma_B=\epsilon\1_2,
\nonumber\\
&\Delta=\begin{pmatrix}
t_x+w_x\delta_x^{-1} & -(t_y+w_y\delta_y) \\
t_y+w_y\delta_y^{-1} & t_x+w_x\delta_x
\end{pmatrix},
\end{alignat}
where $\epsilon$ is an infinitesimal mass term that lifts the degeneracy of the corner states.  
This model is known to exhibit a topological phase for $|t_x/w_x|<1$ and $|t_y/w_y|<1$, whereas it remains trivial otherwise.

First, we validate our formula in Eq.~(\ref{ReaQuaFinite}) for the clean model  
by comparing it with the generalized Resta formula for the quadrupole moment:  
\begin{alignat}{1}
q_{xyR}=-\frac{1}{2\pi} \Im\log\det \Psi^\dagger e^{i\frac{2\pi XY}{N^2}}\Psi.
\label{QuaRes}
\end{alignat}  
In Fig.~\ref{f:quad_fx}, we compare the quadrupole moment given by Eqs.~(\ref{ReaQuaFinite}) and (\ref{QuaRes})  
under both periodic and open boundary conditions.  
Each result is plotted as a function of $t_x$ with fixed $t_y=0.1$ and $w_x=w_y=1$.  
The transition from the topological phase to the trivial phase occurs at $t_x=1$.  
The Resta formula under periodic boundary conditions accurately reproduces both the topological and trivial phases,  
as well as their phase boundary.  
In contrast, our formula also captures both phases well, yielding the expected quantized values  
of the quadrupole moment. However, the phase transition point is slightly shifted.  
This discrepancy arises from finite-size effects: our formula requires larger systems for improved accuracy.  

\begin{figure}[htb]
\begin{center}
\begin{tabular}{c}
\includegraphics[width=0.95\linewidth]{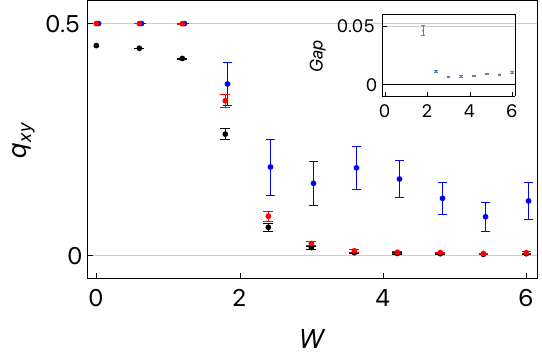}\\
\end{tabular}
\caption{
The quadrupole moment $q_{xy}$ for the disordered BBH model as a function of disorder strength $W$.  
Red, blue, and black points correspond to calculations using Eq.~(\ref{ReaQuaFinite}),  
the Resta formula in Eq.~(\ref{QuaRes}) with periodic boundary conditions,  
and the same Resta formula with open boundary conditions, respectively.  
The parameters are fixed as $t_x=t_y=0.1$, $w_x=w_y=1$, and the system size is $n_0=10$.  
Each point represents an average over 50 disorder realizations.  
The inset shows the bulk energy gap at zero energy.  
}
\label{f:quad_av}
\end{center}
\end{figure}

Now, let us introduce disorder into the BBH model.  
The model includes four hopping parameters, $t_x$, $t_y$, $w_x$, and $w_y$,  
to which we introduce site-dependent randomness as follows:  
$t_a\rightarrow t_a+W\delta t_{a,j}$ and $w_a\rightarrow w_a+W\delta w_{a,j}$ for $a=x,y$,  
where all random variables are independently distributed and defined as  
$t_{a,j}, w_{a,j} \in [-0.5,0.5] + i[-0.5,0.5]$.  

In Fig.~\ref{f:quad_av}, the quadrupole moment calculated using Eq.~(\ref{ReaQuaFinite})  
as a function of disorder strength $W$ is shown, compared with results obtained  
from the Resta formula in Eq.~(\ref{QuaRes}).  
For weak disorder, both our formula and the Resta formula with periodic boundary conditions  
exhibit a nontrivial quantized quadrupole moment.  
As disorder increases, the bulk gap decreases, and the quadrupole moment deviates from the quantized value.  

Our formula, together with the Resta formula under open boundary conditions,  
suggests that a topological transition takes place from the topological quadrupole phase to the trivial phase.  
On the other hand, the Resta formula with periodic boundary conditions exhibits large fluctuations  
even at high disorder strengths. Although the bulk gap appears to reopen,  
our calculations indicate a topological transition, whereas the Resta formula may suggest a gapless phase.  
Further investigation is required to clarify this discrepancy.

\section{Summary and Discussion}\label{s:summary}

We developed a real-space formulation of the Berry phase 
in a manner similar to Kitaev's real-space representation of the Chern number, 
and applied it to various models of topological insulators. 
A well-established real-space method for computing the Berry phase is the Resta formula 
under periodic boundary conditions. 
While this approach successfully reproduces the Berry phase, 
it relies on the assumption of a finite and periodic system, where momentum is discretized. 
In contrast, our method is formulated directly on an infinite lattice, 
providing a rigorous framework for the real-space evaluation of the Berry phase. 

When applying our formulation to concrete problems via numerical calculations, we necessarily use a finite lattice. 
In this case, we employ open boundary conditions, which distinguishes our approach from the Resta formula. 
This difference makes our method complementary to the Resta approach and can be regarded as an alternative perspective on the bulk-edge correspondence, even in systems that are not necessarily topological.
To validate our approach, we examined several disordered models in comparison with the Resta formula 
and obtained consistent results.

We also made an attempt to extend the real-space Berry phase framework to the real-space quadrupole moment, analogous to its extension in the Resta formula. 
Although our formula successfully reproduces the quantized quadrupole moment in the BBH model, 
some inconsistencies arise when compared to the generalized Resta formula for the quadrupole moment under strong disorder.

An intriguing direction for future research is the experimental observation of the Berry phase in real space,  
which could be realized through local measurements of wave-function properties in cold-atom systems.

\acknowledgements
This work was supported in part by a Grant-in-Aid for Scientific Research 
	(Grant No. 22K03448) from the Japan Society for the Promotion of Science.


\begin{thebibliography}{46}
\expandafter\ifx\csname natexlab\endcsname\relax\def\natexlab#1{#1}\fi
\expandafter\ifx\csname bibnamefont\endcsname\relax
  \def\bibnamefont#1{#1}\fi
\expandafter\ifx\csname bibfnamefont\endcsname\relax
  \def\bibfnamefont#1{#1}\fi
\expandafter\ifx\csname citenamefont\endcsname\relax
  \def\citenamefont#1{#1}\fi
\expandafter\ifx\csname url\endcsname\relax
  \def\url#1{\texttt{#1}}\fi
\expandafter\ifx\csname urlprefix\endcsname\relax\def\urlprefix{URL }\fi
\providecommand{\bibinfo}[2]{#2}
\providecommand{\eprint}[2][]{\url{#2}}

\bibitem[{\citenamefont{Kane and Mele}(2005{\natexlab{a}})}]{Kane:2005aa}
\bibinfo{author}{\bibfnamefont{C.~L.} \bibnamefont{Kane}} \bibnamefont{and}
  \bibinfo{author}{\bibfnamefont{E.~J.} \bibnamefont{Mele}},
  \bibinfo{journal}{Physical Review Letters} \textbf{\bibinfo{volume}{95}},
  \bibinfo{pages}{146802} (\bibinfo{year}{2005}{\natexlab{a}}),
  \urlprefix\url{http://link.aps.org/doi/10.1103/PhysRevLett.95.146802}.

\bibitem[{\citenamefont{Kane and Mele}(2005{\natexlab{b}})}]{Kane:2005ab}
\bibinfo{author}{\bibfnamefont{C.~L.} \bibnamefont{Kane}} \bibnamefont{and}
  \bibinfo{author}{\bibfnamefont{E.~J.} \bibnamefont{Mele}},
  \bibinfo{journal}{Physical Review Letters} \textbf{\bibinfo{volume}{95}},
  \bibinfo{pages}{226801} (\bibinfo{year}{2005}{\natexlab{b}}),
  \urlprefix\url{http://link.aps.org/doi/10.1103/PhysRevLett.95.226801}.

\bibitem[{\citenamefont{Qi et~al.}(2008)\citenamefont{Qi, Hughes, and
  Zhang}}]{Qi:2008aa}
\bibinfo{author}{\bibfnamefont{X.-L.} \bibnamefont{Qi}},
  \bibinfo{author}{\bibfnamefont{T.~L.} \bibnamefont{Hughes}},
  \bibnamefont{and} \bibinfo{author}{\bibfnamefont{S.-C.} \bibnamefont{Zhang}},
  \bibinfo{journal}{Physical Review B} \textbf{\bibinfo{volume}{78}},
  \bibinfo{pages}{195424} (\bibinfo{year}{2008}),
  \urlprefix\url{http://link.aps.org/doi/10.1103/PhysRevB.78.195424}.

\bibitem[{\citenamefont{Schnyder et~al.}(2008)\citenamefont{Schnyder, Ryu,
  Furusaki, and Ludwig}}]{Schnyder:2008aa}
\bibinfo{author}{\bibfnamefont{A.~P.} \bibnamefont{Schnyder}},
  \bibinfo{author}{\bibfnamefont{S.}~\bibnamefont{Ryu}},
  \bibinfo{author}{\bibfnamefont{A.}~\bibnamefont{Furusaki}}, \bibnamefont{and}
  \bibinfo{author}{\bibfnamefont{A.~W.~W.} \bibnamefont{Ludwig}},
  \bibinfo{journal}{Physical Review B} \textbf{\bibinfo{volume}{78}},
  \bibinfo{pages}{195125} (\bibinfo{year}{2008}),
  \urlprefix\url{http://link.aps.org/doi/10.1103/PhysRevB.78.195125}.

\bibitem[{\citenamefont{Kitaev}(2009)}]{Kitaev:2009fk}
\bibinfo{author}{\bibfnamefont{A.}~\bibnamefont{Kitaev}},
  \bibinfo{journal}{AIP. Conf. Proc.} \textbf{\bibinfo{volume}{1134}},
  \bibinfo{pages}{22} (\bibinfo{year}{2009}), \eprint{arXiv:0901.2686},
  \urlprefix\url{https://doi.org/10.1063/1.3149495}.

\bibitem[{\citenamefont{Zak}(1989)}]{Zak:1989fk}
\bibinfo{author}{\bibfnamefont{J.}~\bibnamefont{Zak}},
  \bibinfo{journal}{Physical Review Letters} \textbf{\bibinfo{volume}{62}},
  \bibinfo{pages}{2747} (\bibinfo{year}{1989}),
  \urlprefix\url{http://link.aps.org/doi/10.1103/PhysRevLett.62.2747}.

\bibitem[{\citenamefont{King-Smith and Vanderbilt}(1993)}]{King-Smith:1993aa}
\bibinfo{author}{\bibfnamefont{R.~D.} \bibnamefont{King-Smith}}
  \bibnamefont{and}
  \bibinfo{author}{\bibfnamefont{D.}~\bibnamefont{Vanderbilt}},
  \bibinfo{journal}{Physical Review B} \textbf{\bibinfo{volume}{47}},
  \bibinfo{pages}{1651} (\bibinfo{year}{1993}),
  \urlprefix\url{https://link.aps.org/doi/10.1103/PhysRevB.47.1651}.

\bibitem[{\citenamefont{Vanderbilt and King-Smith}(1993)}]{Vanderbilt:1993fk}
\bibinfo{author}{\bibfnamefont{D.}~\bibnamefont{Vanderbilt}} \bibnamefont{and}
  \bibinfo{author}{\bibfnamefont{R.~D.} \bibnamefont{King-Smith}},
  \bibinfo{journal}{Physical Review B} \textbf{\bibinfo{volume}{48}},
  \bibinfo{pages}{4442} (\bibinfo{year}{1993}),
  \urlprefix\url{http://link.aps.org/doi/10.1103/PhysRevB.48.4442}.

\bibitem[{\citenamefont{Resta}(1998)}]{Resta:1998aa}
\bibinfo{author}{\bibfnamefont{R.}~\bibnamefont{Resta}},
  \bibinfo{journal}{Physical Review Letters} \textbf{\bibinfo{volume}{80}},
  \bibinfo{pages}{1800} (\bibinfo{year}{1998}),
  \urlprefix\url{https://link.aps.org/doi/10.1103/PhysRevLett.80.1800}.

\bibitem[{\citenamefont{Kitaev}(2006)}]{Kitaev:2006yg}
\bibinfo{author}{\bibfnamefont{A.}~\bibnamefont{Kitaev}},
  \bibinfo{journal}{Annals of Physics} \textbf{\bibinfo{volume}{321}},
  \bibinfo{pages}{2} (\bibinfo{year}{2006}), \eprint{cond-mat/0506438},
  \urlprefix\url{http://arXiv.org/abs/cond-mat/0506438}.

\bibitem[{\citenamefont{Bianco and Resta}(2011)}]{PhysRevB.84.241106}
\bibinfo{author}{\bibfnamefont{R.}~\bibnamefont{Bianco}} \bibnamefont{and}
  \bibinfo{author}{\bibfnamefont{R.}~\bibnamefont{Resta}},
  \bibinfo{journal}{Phys. Rev. B} \textbf{\bibinfo{volume}{84}},
  \bibinfo{pages}{241106} (\bibinfo{year}{2011}),
  \urlprefix\url{https://link.aps.org/doi/10.1103/PhysRevB.84.241106}.

\bibitem[{\citenamefont{Song and Prodan}(2014)}]{PhysRevB.89.224203}
\bibinfo{author}{\bibfnamefont{J.}~\bibnamefont{Song}} \bibnamefont{and}
  \bibinfo{author}{\bibfnamefont{E.}~\bibnamefont{Prodan}},
  \bibinfo{journal}{Phys. Rev. B} \textbf{\bibinfo{volume}{89}},
  \bibinfo{pages}{224203} (\bibinfo{year}{2014}),
  \urlprefix\url{https://link.aps.org/doi/10.1103/PhysRevB.89.224203}.

\bibitem[{\citenamefont{Lin et~al.}(2021)\citenamefont{Lin, Ke, and
  Lee}}]{PhysRevB.103.224208}
\bibinfo{author}{\bibfnamefont{L.}~\bibnamefont{Lin}},
  \bibinfo{author}{\bibfnamefont{Y.}~\bibnamefont{Ke}}, \bibnamefont{and}
  \bibinfo{author}{\bibfnamefont{C.}~\bibnamefont{Lee}},
  \bibinfo{journal}{Phys. Rev. B} \textbf{\bibinfo{volume}{103}},
  \bibinfo{pages}{224208} (\bibinfo{year}{2021}),
  \urlprefix\url{https://link.aps.org/doi/10.1103/PhysRevB.103.224208}.

\bibitem[{\citenamefont{Hamano and Fukui}(2024)}]{PhysRevB.110.045437}
\bibinfo{author}{\bibfnamefont{F.}~\bibnamefont{Hamano}} \bibnamefont{and}
  \bibinfo{author}{\bibfnamefont{T.}~\bibnamefont{Fukui}},
  \bibinfo{journal}{Phys. Rev. B} \textbf{\bibinfo{volume}{110}},
  \bibinfo{pages}{045437} (\bibinfo{year}{2024}),
  \urlprefix\url{https://link.aps.org/doi/10.1103/PhysRevB.110.045437}.

\bibitem[{\citenamefont{Ryu and Hatsugai}(2002)}]{Ryu:2002fk}
\bibinfo{author}{\bibfnamefont{S.}~\bibnamefont{Ryu}} \bibnamefont{and}
  \bibinfo{author}{\bibfnamefont{Y.}~\bibnamefont{Hatsugai}},
  \bibinfo{journal}{Physical Review Letters} \textbf{\bibinfo{volume}{89}},
  \bibinfo{pages}{077002} (\bibinfo{year}{2002}),
  \urlprefix\url{http://link.aps.org/doi/10.1103/PhysRevLett.89.077002}.

\bibitem[{\citenamefont{Fukui et~al.}(2005)\citenamefont{Fukui, Hatsugai, and
  Suzuki}}]{FHS05}
\bibinfo{author}{\bibfnamefont{T.}~\bibnamefont{Fukui}},
  \bibinfo{author}{\bibfnamefont{Y.}~\bibnamefont{Hatsugai}}, \bibnamefont{and}
  \bibinfo{author}{\bibfnamefont{H.}~\bibnamefont{Suzuki}},
  \bibinfo{journal}{Journal of the Physical Society of Japan}
  \textbf{\bibinfo{volume}{74}}, \bibinfo{pages}{1674} (\bibinfo{year}{2005}),
  \urlprefix\url{http://dx.doi.org/10.1143/JPSJ.74.1674}.

\bibitem[{\citenamefont{Wheeler et~al.}(2019)\citenamefont{Wheeler, Wagner, and
  Hughes}}]{PhysRevB.100.245135}
\bibinfo{author}{\bibfnamefont{W.~A.} \bibnamefont{Wheeler}},
  \bibinfo{author}{\bibfnamefont{L.~K.} \bibnamefont{Wagner}},
  \bibnamefont{and} \bibinfo{author}{\bibfnamefont{T.~L.}
  \bibnamefont{Hughes}}, \bibinfo{journal}{Phys. Rev. B}
  \textbf{\bibinfo{volume}{100}}, \bibinfo{pages}{245135}
  (\bibinfo{year}{2019}),
  \urlprefix\url{https://link.aps.org/doi/10.1103/PhysRevB.100.245135}.

\bibitem[{\citenamefont{Kang et~al.}(2019)\citenamefont{Kang, Shiozaki, and
  Cho}}]{PhysRevB.100.245134}
\bibinfo{author}{\bibfnamefont{B.}~\bibnamefont{Kang}},
  \bibinfo{author}{\bibfnamefont{K.}~\bibnamefont{Shiozaki}}, \bibnamefont{and}
  \bibinfo{author}{\bibfnamefont{G.~Y.} \bibnamefont{Cho}},
  \bibinfo{journal}{Phys. Rev. B} \textbf{\bibinfo{volume}{100}},
  \bibinfo{pages}{245134} (\bibinfo{year}{2019}),
  \urlprefix\url{https://link.aps.org/doi/10.1103/PhysRevB.100.245134}.

\bibitem[{\citenamefont{Li et~al.}(2020)\citenamefont{Li, Fu, Hu, Li, and
  Shen}}]{PhysRevLett.125.166801}
\bibinfo{author}{\bibfnamefont{C.-A.} \bibnamefont{Li}},
  \bibinfo{author}{\bibfnamefont{B.}~\bibnamefont{Fu}},
  \bibinfo{author}{\bibfnamefont{Z.-A.} \bibnamefont{Hu}},
  \bibinfo{author}{\bibfnamefont{J.}~\bibnamefont{Li}}, \bibnamefont{and}
  \bibinfo{author}{\bibfnamefont{S.-Q.} \bibnamefont{Shen}},
  \bibinfo{journal}{Phys. Rev. Lett.} \textbf{\bibinfo{volume}{125}},
  \bibinfo{pages}{166801} (\bibinfo{year}{2020}),
  \urlprefix\url{https://link.aps.org/doi/10.1103/PhysRevLett.125.166801}.

\bibitem[{\citenamefont{Benalcazar and Cerjan}(2022)}]{PhysRevLett.128.127601}
\bibinfo{author}{\bibfnamefont{W.~A.} \bibnamefont{Benalcazar}}
  \bibnamefont{and} \bibinfo{author}{\bibfnamefont{A.}~\bibnamefont{Cerjan}},
  \bibinfo{journal}{Phys. Rev. Lett.} \textbf{\bibinfo{volume}{128}},
  \bibinfo{pages}{127601} (\bibinfo{year}{2022}),
  \urlprefix\url{https://link.aps.org/doi/10.1103/PhysRevLett.128.127601}.

\bibitem[{\citenamefont{Benalcazar
  et~al.}(2017{\natexlab{a}})\citenamefont{Benalcazar, Bernevig, and
  Hughes}}]{Benalcazar:2017aa}
\bibinfo{author}{\bibfnamefont{W.~A.} \bibnamefont{Benalcazar}},
  \bibinfo{author}{\bibfnamefont{B.~A.} \bibnamefont{Bernevig}},
  \bibnamefont{and} \bibinfo{author}{\bibfnamefont{T.~L.}
  \bibnamefont{Hughes}}, \bibinfo{journal}{Physical Review B}
  \textbf{\bibinfo{volume}{96}}, \bibinfo{pages}{245115}
  (\bibinfo{year}{2017}{\natexlab{a}}).

\bibitem[{\citenamefont{Benalcazar
  et~al.}(2017{\natexlab{b}})\citenamefont{Benalcazar, Bernevig, and
  Hughes}}]{Benalcazar:2017ab}
\bibinfo{author}{\bibfnamefont{W.~A.} \bibnamefont{Benalcazar}},
  \bibinfo{author}{\bibfnamefont{B.~A.} \bibnamefont{Bernevig}},
  \bibnamefont{and} \bibinfo{author}{\bibfnamefont{T.~L.}
  \bibnamefont{Hughes}}, \bibinfo{journal}{Science}
  \textbf{\bibinfo{volume}{357}}, \bibinfo{pages}{61}
  (\bibinfo{year}{2017}{\natexlab{b}}).

\bibitem[{\citenamefont{Schindler
  et~al.}(2018{\natexlab{a}})\citenamefont{Schindler, Cook, Vergniory, Wang,
  Parkin, Bernevig, and Neupert}}]{Schindler:2018ab}
\bibinfo{author}{\bibfnamefont{F.}~\bibnamefont{Schindler}},
  \bibinfo{author}{\bibfnamefont{A.~M.} \bibnamefont{Cook}},
  \bibinfo{author}{\bibfnamefont{M.~G.} \bibnamefont{Vergniory}},
  \bibinfo{author}{\bibfnamefont{Z.}~\bibnamefont{Wang}},
  \bibinfo{author}{\bibfnamefont{S.~S.~P.} \bibnamefont{Parkin}},
  \bibinfo{author}{\bibfnamefont{B.~A.} \bibnamefont{Bernevig}},
  \bibnamefont{and} \bibinfo{author}{\bibfnamefont{T.}~\bibnamefont{Neupert}},
  \bibinfo{journal}{Science Advances} \textbf{\bibinfo{volume}{4}},
  \bibinfo{pages}{eaat0346} (\bibinfo{year}{2018}{\natexlab{a}}),
  \urlprefix\url{https://advances.sciencemag.org/content/advances/4/6/eaat0346.full.pdf}.

\bibitem[{\citenamefont{Imhof et~al.}(2018)\citenamefont{Imhof, Berger, Bayer,
  Brehm, Molenkamp, Kiessling, Schindler, Lee, Greiter, Neupert
  et~al.}}]{Imhof:2018aa}
\bibinfo{author}{\bibfnamefont{S.}~\bibnamefont{Imhof}},
  \bibinfo{author}{\bibfnamefont{C.}~\bibnamefont{Berger}},
  \bibinfo{author}{\bibfnamefont{F.}~\bibnamefont{Bayer}},
  \bibinfo{author}{\bibfnamefont{J.}~\bibnamefont{Brehm}},
  \bibinfo{author}{\bibfnamefont{L.~W.} \bibnamefont{Molenkamp}},
  \bibinfo{author}{\bibfnamefont{T.}~\bibnamefont{Kiessling}},
  \bibinfo{author}{\bibfnamefont{F.}~\bibnamefont{Schindler}},
  \bibinfo{author}{\bibfnamefont{C.~H.} \bibnamefont{Lee}},
  \bibinfo{author}{\bibfnamefont{M.}~\bibnamefont{Greiter}},
  \bibinfo{author}{\bibfnamefont{T.}~\bibnamefont{Neupert}},
  \bibnamefont{et~al.}, \bibinfo{journal}{Nature Physics}
  \textbf{\bibinfo{volume}{14}}, \bibinfo{pages}{925} (\bibinfo{year}{2018}).

\bibitem[{\citenamefont{Kunst et~al.}(2018)\citenamefont{Kunst, van Miert, and
  Bergholtz}}]{Kunst:2018aa}
\bibinfo{author}{\bibfnamefont{F.~K.} \bibnamefont{Kunst}},
  \bibinfo{author}{\bibfnamefont{G.}~\bibnamefont{van Miert}},
  \bibnamefont{and} \bibinfo{author}{\bibfnamefont{E.~J.}
  \bibnamefont{Bergholtz}}, \bibinfo{journal}{Physical Review B}
  \textbf{\bibinfo{volume}{97}}, \bibinfo{pages}{241405}
  (\bibinfo{year}{2018}),
  \urlprefix\url{https://link.aps.org/doi/10.1103/PhysRevB.97.241405}.

\bibitem[{\citenamefont{Ezawa}(2018)}]{Ezawa:2018aa}
\bibinfo{author}{\bibfnamefont{M.}~\bibnamefont{Ezawa}},
  \bibinfo{journal}{Physical Review Letters} \textbf{\bibinfo{volume}{120}},
  \bibinfo{pages}{026801} (\bibinfo{year}{2018}).

\bibitem[{\citenamefont{Schindler
  et~al.}(2018{\natexlab{b}})\citenamefont{Schindler, Cook, Vergniory, Wang,
  Parkin, Bernevig, and Neupert}}]{Schindler:2018aa}
\bibinfo{author}{\bibfnamefont{F.}~\bibnamefont{Schindler}},
  \bibinfo{author}{\bibfnamefont{A.~M.} \bibnamefont{Cook}},
  \bibinfo{author}{\bibfnamefont{M.~G.} \bibnamefont{Vergniory}},
  \bibinfo{author}{\bibfnamefont{Z.}~\bibnamefont{Wang}},
  \bibinfo{author}{\bibfnamefont{S.~S.~P.} \bibnamefont{Parkin}},
  \bibinfo{author}{\bibfnamefont{B.~A.} \bibnamefont{Bernevig}},
  \bibnamefont{and} \bibinfo{author}{\bibfnamefont{T.}~\bibnamefont{Neupert}},
  \bibinfo{journal}{Science Advances} \textbf{\bibinfo{volume}{4}}
  (\bibinfo{year}{2018}{\natexlab{b}}).

\bibitem[{\citenamefont{Fukui and Hatsugai}(2018)}]{Fukui:2018aa}
\bibinfo{author}{\bibfnamefont{T.}~\bibnamefont{Fukui}} \bibnamefont{and}
  \bibinfo{author}{\bibfnamefont{Y.}~\bibnamefont{Hatsugai}},
  \bibinfo{journal}{Physical Review B} \textbf{\bibinfo{volume}{98}},
  \bibinfo{pages}{035147} (\bibinfo{year}{2018}).

\bibitem[{\citenamefont{Hayashi}(2018)}]{Hayashi:2018aa}
\bibinfo{author}{\bibfnamefont{S.}~\bibnamefont{Hayashi}},
  \bibinfo{journal}{Communications in Mathematical Physics}
  \textbf{\bibinfo{volume}{364}}, \bibinfo{pages}{343} (\bibinfo{year}{2018}).

\bibitem[{\citenamefont{Hashimoto et~al.}(2017)\citenamefont{Hashimoto, Wu, and
  Kimura}}]{Hashimoto:2017aa}
\bibinfo{author}{\bibfnamefont{K.}~\bibnamefont{Hashimoto}},
  \bibinfo{author}{\bibfnamefont{X.}~\bibnamefont{Wu}}, \bibnamefont{and}
  \bibinfo{author}{\bibfnamefont{T.}~\bibnamefont{Kimura}},
  \bibinfo{journal}{Physical Review B} \textbf{\bibinfo{volume}{95}},
  \bibinfo{pages}{165443} (\bibinfo{year}{2017}).

\bibitem[{\citenamefont{Ota et~al.}(2019)\citenamefont{Ota, Liu, Katsumi,
  Watanabe, Wakabayashi, Arakawa, and Iwamoto}}]{Ota:2019aa}
\bibinfo{author}{\bibfnamefont{Y.}~\bibnamefont{Ota}},
  \bibinfo{author}{\bibfnamefont{F.}~\bibnamefont{Liu}},
  \bibinfo{author}{\bibfnamefont{R.}~\bibnamefont{Katsumi}},
  \bibinfo{author}{\bibfnamefont{K.}~\bibnamefont{Watanabe}},
  \bibinfo{author}{\bibfnamefont{K.}~\bibnamefont{Wakabayashi}},
  \bibinfo{author}{\bibfnamefont{Y.}~\bibnamefont{Arakawa}}, \bibnamefont{and}
  \bibinfo{author}{\bibfnamefont{S.}~\bibnamefont{Iwamoto}},
  \bibinfo{journal}{Optica} \textbf{\bibinfo{volume}{6}}, \bibinfo{pages}{786}
  (\bibinfo{year}{2019}).

\bibitem[{\citenamefont{Zhang et~al.}(2019)\citenamefont{Zhang, Wang, Lin,
  Tian, Xie, Lu, Chen, and Jiang}}]{Zhang:2019ab}
\bibinfo{author}{\bibfnamefont{X.}~\bibnamefont{Zhang}},
  \bibinfo{author}{\bibfnamefont{H.-X.} \bibnamefont{Wang}},
  \bibinfo{author}{\bibfnamefont{Z.-K.} \bibnamefont{Lin}},
  \bibinfo{author}{\bibfnamefont{Y.}~\bibnamefont{Tian}},
  \bibinfo{author}{\bibfnamefont{B.}~\bibnamefont{Xie}},
  \bibinfo{author}{\bibfnamefont{M.-H.} \bibnamefont{Lu}},
  \bibinfo{author}{\bibfnamefont{Y.-F.} \bibnamefont{Chen}}, \bibnamefont{and}
  \bibinfo{author}{\bibfnamefont{J.-H.} \bibnamefont{Jiang}},
  \bibinfo{journal}{Nature Physics} \textbf{\bibinfo{volume}{15}},
  \bibinfo{pages}{582} (\bibinfo{year}{2019}).

\bibitem[{\citenamefont{Hayward et~al.}(2021)\citenamefont{Hayward, Bertok,
  Schneider, and Heidrich-Meisner}}]{PhysRevA.103.043310}
\bibinfo{author}{\bibfnamefont{A.~L.~C.} \bibnamefont{Hayward}},
  \bibinfo{author}{\bibfnamefont{E.}~\bibnamefont{Bertok}},
  \bibinfo{author}{\bibfnamefont{U.}~\bibnamefont{Schneider}},
  \bibnamefont{and}
  \bibinfo{author}{\bibfnamefont{F.}~\bibnamefont{Heidrich-Meisner}},
  \bibinfo{journal}{Phys. Rev. A} \textbf{\bibinfo{volume}{103}},
  \bibinfo{pages}{043310} (\bibinfo{year}{2021}),
  \urlprefix\url{https://link.aps.org/doi/10.1103/PhysRevA.103.043310}.

\bibitem[{\citenamefont{Thouless}(1983)}]{Thouless:1983fk}
\bibinfo{author}{\bibfnamefont{D.~J.} \bibnamefont{Thouless}},
  \bibinfo{journal}{Physical Review B} \textbf{\bibinfo{volume}{27}},
  \bibinfo{pages}{6083} (\bibinfo{year}{1983}),
  \urlprefix\url{http://link.aps.org/doi/10.1103/PhysRevB.27.6083}.

\bibitem[{\citenamefont{Lohse et~al.}(2016)\citenamefont{Lohse, Schweizer,
  Zilberberg, Aidelsburger, and Bloch}}]{Lohse:2016aa}
\bibinfo{author}{\bibfnamefont{M.}~\bibnamefont{Lohse}},
  \bibinfo{author}{\bibfnamefont{C.}~\bibnamefont{Schweizer}},
  \bibinfo{author}{\bibfnamefont{O.}~\bibnamefont{Zilberberg}},
  \bibinfo{author}{\bibfnamefont{M.}~\bibnamefont{Aidelsburger}},
  \bibnamefont{and} \bibinfo{author}{\bibfnamefont{I.}~\bibnamefont{Bloch}},
  \bibinfo{journal}{Nat Phys} \textbf{\bibinfo{volume}{12}},
  \bibinfo{pages}{350} (\bibinfo{year}{2016}),
  \urlprefix\url{http://dx.doi.org/10.1038/nphys3584}.

\bibitem[{\citenamefont{Nakajima et~al.}(2016)\citenamefont{Nakajima, Tomita,
  Taie, Ichinose, Ozawa, Wang, Troyer, and Takahashi}}]{Nakajima:2016aa}
\bibinfo{author}{\bibfnamefont{S.}~\bibnamefont{Nakajima}},
  \bibinfo{author}{\bibfnamefont{T.}~\bibnamefont{Tomita}},
  \bibinfo{author}{\bibfnamefont{S.}~\bibnamefont{Taie}},
  \bibinfo{author}{\bibfnamefont{T.}~\bibnamefont{Ichinose}},
  \bibinfo{author}{\bibfnamefont{H.}~\bibnamefont{Ozawa}},
  \bibinfo{author}{\bibfnamefont{L.}~\bibnamefont{Wang}},
  \bibinfo{author}{\bibfnamefont{M.}~\bibnamefont{Troyer}}, \bibnamefont{and}
  \bibinfo{author}{\bibfnamefont{Y.}~\bibnamefont{Takahashi}},
  \bibinfo{journal}{Nat Phys} \textbf{\bibinfo{volume}{12}},
  \bibinfo{pages}{296} (\bibinfo{year}{2016}),
  \urlprefix\url{http://dx.doi.org/10.1038/nphys3622}.

\bibitem[{\citenamefont{Wang et~al.}(2013)\citenamefont{Wang, Troyer, and
  Dai}}]{Wang:2013fk_pump}
\bibinfo{author}{\bibfnamefont{L.}~\bibnamefont{Wang}},
  \bibinfo{author}{\bibfnamefont{M.}~\bibnamefont{Troyer}}, \bibnamefont{and}
  \bibinfo{author}{\bibfnamefont{X.}~\bibnamefont{Dai}},
  \bibinfo{journal}{Physical Review Letters} \textbf{\bibinfo{volume}{111}},
  \bibinfo{pages}{026802} (\bibinfo{year}{2013}),
  \urlprefix\url{http://link.aps.org/doi/10.1103/PhysRevLett.111.026802}.

\bibitem[{\citenamefont{Prodan}(2009)}]{PhysRevB.80.125327}
\bibinfo{author}{\bibfnamefont{E.}~\bibnamefont{Prodan}},
  \bibinfo{journal}{Phys. Rev. B} \textbf{\bibinfo{volume}{80}},
  \bibinfo{pages}{125327} (\bibinfo{year}{2009}),
  \urlprefix\url{https://link.aps.org/doi/10.1103/PhysRevB.80.125327}.

\bibitem[{\citenamefont{Prodan}(2010)}]{Prodan_2010}
\bibinfo{author}{\bibfnamefont{E.}~\bibnamefont{Prodan}}, \bibinfo{journal}{New
  Journal of Physics} \textbf{\bibinfo{volume}{12}}, \bibinfo{pages}{065003}
  (\bibinfo{year}{2010}),
  \urlprefix\url{https://dx.doi.org/10.1088/1367-2630/12/6/065003}.

\bibitem[{\citenamefont{Caio et~al.}(2019)\citenamefont{Caio, M{\"o}ller,
  Cooper, and Bhaseen}}]{Caio:2019aa}
\bibinfo{author}{\bibfnamefont{M.~D.} \bibnamefont{Caio}},
  \bibinfo{author}{\bibfnamefont{G.}~\bibnamefont{M{\"o}ller}},
  \bibinfo{author}{\bibfnamefont{N.~R.} \bibnamefont{Cooper}},
  \bibnamefont{and} \bibinfo{author}{\bibfnamefont{M.~J.}
  \bibnamefont{Bhaseen}}, \bibinfo{journal}{Nature Physics}
  \textbf{\bibinfo{volume}{15}}, \bibinfo{pages}{257} (\bibinfo{year}{2019}),
  \urlprefix\url{https://doi.org/10.1038/s41567-018-0390-7}.

\bibitem[{\citenamefont{Irsigler et~al.}(2019)\citenamefont{Irsigler, Zheng,
  and Hofstetter}}]{PhysRevA.100.023610}
\bibinfo{author}{\bibfnamefont{B.}~\bibnamefont{Irsigler}},
  \bibinfo{author}{\bibfnamefont{J.-H.} \bibnamefont{Zheng}}, \bibnamefont{and}
  \bibinfo{author}{\bibfnamefont{W.}~\bibnamefont{Hofstetter}},
  \bibinfo{journal}{Phys. Rev. A} \textbf{\bibinfo{volume}{100}},
  \bibinfo{pages}{023610} (\bibinfo{year}{2019}),
  \urlprefix\url{https://link.aps.org/doi/10.1103/PhysRevA.100.023610}.

\bibitem[{\citenamefont{Marrazzo and Resta}(2019)}]{PhysRevLett.122.166602}
\bibinfo{author}{\bibfnamefont{A.}~\bibnamefont{Marrazzo}} \bibnamefont{and}
  \bibinfo{author}{\bibfnamefont{R.}~\bibnamefont{Resta}},
  \bibinfo{journal}{Phys. Rev. Lett.} \textbf{\bibinfo{volume}{122}},
  \bibinfo{pages}{166602} (\bibinfo{year}{2019}),
  \urlprefix\url{https://link.aps.org/doi/10.1103/PhysRevLett.122.166602}.

\bibitem[{\citenamefont{Gebert et~al.}(2020)\citenamefont{Gebert, Irsigler, and
  Hofstetter}}]{PhysRevA.101.063606}
\bibinfo{author}{\bibfnamefont{U.}~\bibnamefont{Gebert}},
  \bibinfo{author}{\bibfnamefont{B.}~\bibnamefont{Irsigler}}, \bibnamefont{and}
  \bibinfo{author}{\bibfnamefont{W.}~\bibnamefont{Hofstetter}},
  \bibinfo{journal}{Phys. Rev. A} \textbf{\bibinfo{volume}{101}},
  \bibinfo{pages}{063606} (\bibinfo{year}{2020}),
  \urlprefix\url{https://link.aps.org/doi/10.1103/PhysRevA.101.063606}.

\bibitem[{\citenamefont{Sykes and Barnett}(2021)}]{PhysRevB.103.155134}
\bibinfo{author}{\bibfnamefont{J.}~\bibnamefont{Sykes}} \bibnamefont{and}
  \bibinfo{author}{\bibfnamefont{R.}~\bibnamefont{Barnett}},
  \bibinfo{journal}{Phys. Rev. B} \textbf{\bibinfo{volume}{103}},
  \bibinfo{pages}{155134} (\bibinfo{year}{2021}),
  \urlprefix\url{https://link.aps.org/doi/10.1103/PhysRevB.103.155134}.

\bibitem[{\citenamefont{Shiina et~al.}(2025)\citenamefont{Shiina, Hamano, and
  Fukui}}]{shiina2025realspacerepresentationsecondchern}
\bibinfo{author}{\bibfnamefont{T.}~\bibnamefont{Shiina}},
  \bibinfo{author}{\bibfnamefont{F.}~\bibnamefont{Hamano}}, \bibnamefont{and}
  \bibinfo{author}{\bibfnamefont{T.}~\bibnamefont{Fukui}},
  \emph{\bibinfo{title}{Real-space representation of the second chern number}}
  (\bibinfo{year}{2025}), \eprint{2502.14299},
  \urlprefix\url{https://arxiv.org/abs/2502.14299}.

\bibitem[{\citenamefont{Hatsugai and Fukui}(2016)}]{Hatsugai:2016aa}
\bibinfo{author}{\bibfnamefont{Y.}~\bibnamefont{Hatsugai}} \bibnamefont{and}
  \bibinfo{author}{\bibfnamefont{T.}~\bibnamefont{Fukui}},
  \bibinfo{journal}{Physical Review B} \textbf{\bibinfo{volume}{94}},
  \bibinfo{pages}{041102} (\bibinfo{year}{2016}),
  \urlprefix\url{http://link.aps.org/doi/10.1103/PhysRevB.94.041102}.

\end{thebibliography}

\end{document}